\documentclass[twocolumn,american,english]{revtex4-2}
\usepackage[T1]{fontenc}
\usepackage[latin9]{luainputenc}
\usepackage{amsmath}
\usepackage{amssymb}
\usepackage{babel}
\begin{document}
\title{Order parameters in quasi-1D spin systems}
\author{Garry Goldstein$^{1}$}
\address{$^{1}$garrygoldsteinwinnipeg@gmail.com}
\begin{abstract}
In this work we extend the notion of what is meant by a meanfield.
For the purposes of this work meanfields are approximately maps -
through some self consistency relation - of a complex, usually manybody,
problem to a simpler more readily solvable problem. This mapping can
then be solved to represent properties of the complex many body problem
using some self consistency relations and the solution of the simpler
problem. Prototypical examples of simpler meanfield problems (meanfield
systems) are the single site and free particle problems - which are
exactly solvable. Here we propose a new class of simple meanfield
systems where the simple problem to be solved is a 1D spin chain.
These meanfields are particularly useful for studying quasi-1D models,
where there is a 3D system composed of weakly coupled 1D spin chains
with the coupling in the transverse direction weaker then in the 1D
direction. We illustrate this idea by considering meanfields for the
Ising (of any coupling sign) and ferromagnetic Heisenberg models with
one direction coupled much more strongly then the other directions
(quasi-1D systems) which map at meanfield level onto the 1D Ising
and 1D ferromagnetic Heisenberg models. We also consider more exotic
models to illustrate other methods of solving 1D systems, namely the
ferromagnetic $N$-state Potts model. Magnetic phase transition temperatures
and are obtained for all three models, we see that they significantly
differ from the usual meanfield estimates. Indeed if the 1D direction
has coupling $\Gamma$ and the transverse directions have coupling
$J$ with $\lambda\sim\frac{\Gamma}{J}\gg1$ then regular meanfield
would predict the transition temperature to be $k_{B}T_{c}\sim\Gamma$
for all three models while 1D meanfield predicts temperatures of $k_{B}T_{c}\sim\frac{\Gamma}{\log\left(\lambda\right)}$
for the Ising and Potts models and $k_{B}T_{c}\sim\frac{\Gamma}{\sqrt{\lambda}}$
for the ferromagnetic Heisenberg model. Cluster 1D (ladders etc.)
meanfield extensions are also proposed.
\end{abstract}
\maketitle

\section{\protect\label{sec:Main-idea}Introduction}

Meanfields have been very successfully and almost universally used
as a first step towards the solution (understanding) of many complex
manybody problems \citep{Pathria_2011,Sachdev_2011,Sachdev_2023,Coleman_2016,Continentino_2017}.
In a meanfield like solution of a complex problem we map a difficult
manybody problem into a simpler one (a simpler system which can then
be efficiently solved), solve the problem and solve for self consistency
to relate the parameters of the solution of the simpler problem to
the more complex many body problem to be solved or understood. The
prototypical meanfield - simple - systems are the single site problem
or the single particle problem. In the single particle case the complex
system is mapped onto a quadratic Hamiltonian, through say a Hartree-Fock
\citep{Coleman_2016} meanfield, in a self consistent manner whereby
properties of the complex model namely ground state energy, symmetry
breaking, correlation functions may be often reliably computed to
some accuracy from the Hartree-Fock problem \citep{Coleman_2016}.
In the single site problem - often a spin model is mapped onto a single
spin problem which can be efficiently solved. Then self consistency
is imposed through equating the parameters of the single site problem
to the effects of the couplings to the neighboring spins and their
magnetization \citep{Pathria_2011}. In some cases, the single site
problem may be extended, as in Dynamical Meanfield Theory (DMFT),
to a single site and a bath whereby the frequency dependence (but
not momentum dependence) of the Green's functions of the manybody
system may be modeled self consistently through the dissipitative
effects of the bath \citep{Georges_1996}. 

Here we propose another important simple system which may be effciently
analyzed as a meanfield system - the 1D system. Indeed many 1D models
are solvable through transfer matrix calculations \citep{Pathria_2011,Baxter_1982},
Density Matrix Renormalization Group (DMRG) \citep{Schollwock_2005,Verstraete_2023,Catarina_2023}
methods , Jordan-Wigner fermionization \citep{Sachdev_2011,Coleman_2016},
or more generally the Bethe Ansatz (BA) methods \citep{Takahashi_1999,Zvyagin_2005,Baxter_1982,Andrei_1994},
and Exact Diagonalization (ED) techniques \citep{Wiebe_2008,Jung_2020,Tubman_2020}
to name a few. In this work we consider 3D spin systems where the
coupling in one direction is much stronger then the coupling in the
other two (quasi-1D systems). For these systems mapping the system
to a 1D meanfield system and solving for self consistency is more
efficient then regular meanfield (we verify this in part in Appendix
\ref{sec:Onsanger-solution}). We illustrate this idea through the
3D (quasi-1D) Ising model whose meanfield is the 1D Ising model (of
either sign of coupling) in an external field (which is solvable by
transfer matrix techniques) and the 3D (quasi-1D) ferromagnetic Heisenberg
model whose meanfield is the 1D ferromagnetic Heisenberg model in
an external field (which is solvable by BA techniques). We choose
these two examples because of their simplicity clarity and because
in both 3D and 1D these models are prototypical examples of models
with magnetic phase transitions. We also illustrate the ideas through
a more exotic related example the $N$-state ferromagnetic Potts model.
We show significant differences between regular meanfield results
and 1D meanfield results. Indeed assuming two couplings $\Gamma\gg J$
for longitudinal and transverse directions and $\lambda\sim\frac{\Gamma}{J}\gg1$;
then regular meanfield predicts a transition temperature of $k_{B}T_{c}\sim\Gamma$
for all three models while 1D improved meanfield shows that $k_{B}T_{c}\sim\frac{\Gamma}{\log\left(\lambda\right)}$
for the Ising and Potts models and $k_{B}T_{c}\sim\frac{\Gamma}{\sqrt{\lambda}}$
for the ferromagnetic Heisenberg model. Here $k_{B}$ is the Boltzmann
constant and $T_{c}$ is the critical temperature for the magnetic
phase transition. Cluster 1D meanfields are also proposed, which should
systematically improve meanfield accuracy \citep{Pathria_2011}.

\section{\protect\label{sec:Ising-exmple}Ising model}

Consider a 3D quasi-1D ferromagnetic Ising model in an external field
with the following Hamiltonian: 
\begin{align}
H_{I}= & -\Gamma\sum_{i}\sigma^{z}\left(i\right)\sigma^{z}\left(i+\hat{z}\right)\nonumber \\
 & -J\sum_{\left\langle i,j\right\rangle ,j\neq i\pm\hat{z}}\sigma^{z}\left(i\right)\sigma^{z}\left(j\right)+B\sum_{i}\sigma^{z}\left(i\right)\label{eq:Ising_external_field}
\end{align}
at a temperature $\beta$ and $\left\langle i,j\right\rangle $ are
nearest neighbors, here we will be interested in $\Gamma\gg J$ so
that the 1D chains along the z-axis are more strongly coupled then
the transverse plane. Antiferromagnetic Ising models are equivalent
to ferromagnetic Ising models on bipartite lattices. so need not be
considered Now consider the following meanfield Hamiltonian at unit
temperature: 

\begin{equation}
H_{MF}=\gamma\sum_{i}^{N}\sigma^{z}\left(i\right)\sigma^{z}\left(i+1\right)+h\sum_{i}\sigma^{z}\left(i\right)\label{eq:Hamiltonian}
\end{equation}
Where for self consistency:
\begin{equation}
\gamma=\beta\Gamma,\;h=\beta J\mathcal{N}m-\beta B\label{eq:Meanfield}
\end{equation}
here $\mathcal{N}\sim2\left(d-1\right)$ is the number of nearest
neighbors in the weakly coupled directions and $d=3$ is the dimension
of the system. Then we know that the magnetization $m$ in the thermodynamic
limit for the Hamiltonian in Eq. (\ref{eq:Hamiltonian}) at unit temperature
is given by \citet{Pathria_2011}:
\begin{equation}
m=\frac{\sinh\left(h\right)}{\left[\exp\left(-4\gamma\right)+\left(\sinh\left(h\right)\right)^{2}\right]^{1/2}}\label{eq:Magnetization}
\end{equation}
This is an exact result (using the transfer matrix formalism) \citep{Pathria_2011,Baxter_1982}.
Now we substitute the meanfield relations in Eq. (\ref{eq:Meanfield})
and obtain: 
\begin{equation}
m=\frac{\sinh\left(\beta J\mathcal{N}m-\beta B\right)}{\left[\exp\left(-4\beta\Gamma\right)+\left(\sinh\left(\beta J\mathcal{N}m-\beta B\right)\right)^{2}\right]^{1/2}}\label{eq:Magnetization-1}
\end{equation}
Setting $B=0$ and linearizing to find the phase transition temperature
we have that:
\begin{align}
m & =\exp\left(2\beta\Gamma\right)\beta J\mathcal{N}m,\nonumber \\
1 & =\exp\left(2\beta\Gamma\right)\beta J\mathcal{N}\label{eq:Magnetism}
\end{align}
leads to magnetism. Taking the logarithm of both sides we see that:
\begin{equation}
2\beta\Gamma=-\log\left(\beta J\mathcal{N}\right)\label{eq:Critical_temp}
\end{equation}
Introducing 
\begin{equation}
\lambda=\frac{\Gamma}{J\mathcal{N}}\gg1\label{eq:Lambda}
\end{equation}
We get that: 
\begin{equation}
2\lambda=-\frac{\log\left(\beta J\mathcal{N}\right)}{\beta J\mathcal{N}}\label{eq:Transition-2}
\end{equation}
We now write: 
\begin{equation}
\beta_{0}J\mathcal{N}=\frac{1}{2\lambda}\label{eq:Iterate-1}
\end{equation}
Then: 
\begin{equation}
-\frac{\log\left(\beta J\mathcal{N}\right)}{\beta J\mathcal{N}}\cong\frac{\log\left(2\lambda\right)}{\beta J\mathcal{N}}\label{eq:First_approx}
\end{equation}
This means that: 
\begin{equation}
\beta_{1}J\mathcal{N}=\frac{1}{2\lambda\log\left(2\lambda\right)}\label{eq:Solution-1}
\end{equation}
In general we can iterate the solution through the relation (though
Eq. (\ref{eq:Solution-1}) is often enough): 
\begin{equation}
\beta_{n+1}J\mathcal{N}=-\frac{1}{2\lambda\log\left(\beta_{n}J\mathcal{N}\right)}\label{eq:Iterate-2}
\end{equation}
We note that the usual meanfield equations the phase transition temperature
can be found through the following relationship:
\begin{equation}
2\beta\Gamma+\beta J\mathcal{N}=1\label{eq:Worse}
\end{equation}
Which are much worse as it predicts:
\begin{equation}
\beta J\mathcal{N}=\frac{1}{2\lambda+1}\label{eq:Lowest_order}
\end{equation}
which is significantly different, for large $\lambda$, then Eq. (\ref{eq:Solution-1}),
see Appendix \ref{sec:Onsanger-solution}.

\section{\protect\label{sec:-state-Potts}$N$ state Potts model}

We focus on the ferromagnetic Potts model. We write: 
\begin{align}
H_{Pott}= & -\Gamma\sum_{i}\delta\left(n\left(i\right),n\left(i+\hat{z}\right)\right)\nonumber \\
 & -J\sum_{\left\langle i,j\right\rangle ,j\neq i\pm\hat{z}}\delta\left(n\left(i\right),n\left(j\right)\right)+B\sum_{i}\delta\left(n\left(i\right),1\right)\label{eq:Potts_model}
\end{align}
Where $n\left(i\right)$ is the state of the i'th unit, with $n\left(i\right)=1,2...,N$.
We now write the Hamiltonian of the meanfield system: 
\begin{equation}
H_{Pott}=-\Gamma\sum_{i}\delta\left(n\left(i\right),n\left(i+1\right)\right)-\mathcal{B}\sum_{i}\delta\left(n\left(i\right),1\right)\label{eq:Potts}
\end{equation}
Then we write the transfer matrix \citep{Baxter_1982,Pathria_2011}
of the system as:
\begin{align}
T & =\left|V\right\rangle \left\langle V\right|+\left(\exp\left(\mathcal{B}\right)-1\right)\left[\left|V\right\rangle \left\langle 1\right|+\left|1\right\rangle \left\langle V\right|\right]\nonumber \\
 & \quad+\left(\exp\left(\Gamma\right)-1\right)\sum_{i=2}^{N}\left|i\right\rangle \left\langle i\right|+\nonumber \\
 & \quad+\left(\exp\left(\Gamma+2\mathcal{B}\right)-2\exp\left(\mathcal{B}\right)+1\right)\left|1\right\rangle \left\langle 1\right|\label{eq:Transfer}
\end{align}
 Where we introduce the un-normalized vector 
\begin{equation}
\left|V\right\rangle =\sum_{i=1}^{N}\left|i\right\rangle \label{eq:Vector-1}
\end{equation}
Now we look for the biggest eigenvalue of the transfer matrix \citep{Baxter_1982,Pathria_2011}
using the ansatz: 
\begin{align}
 & E\left(a\left|1\right\rangle +b\left|V\right\rangle \right)\nonumber \\
 & =T\left(a\left|1\right\rangle +b\left|V\right\rangle \right)\label{eq:Diagonal}
\end{align}

\begin{widetext}
Now we linearIze everything for small $\mathfrak{B}$ in order to
find the magnetic phase transition and write:

\begin{equation}
\left(\begin{array}{cc}
\left(\exp\left(\beta\Gamma\right)-\left(2\exp\left(\beta\Gamma\right)-1\right)\beta\mathcal{B}\right) & \left(\left(N-2\right)+2\exp\left(\beta\Gamma\right)\right)\beta\mathcal{B}\\
\beta\mathcal{B} & \left(N-1+\exp\left(\beta\Gamma\right)+\beta\mathcal{B}\right)
\end{array}\right)\left(\begin{array}{c}
a\\
b
\end{array}\right)=E\left(\begin{array}{c}
a\\
b
\end{array}\right)\label{eq:Eigenvalue-1}
\end{equation}
For a full calculation to all orders see Appendix \ref{sec:-state-Potts-model}.
Now we treat the matrix proportional to $\beta\mathcal{B}$ as a perturbation
and we write:
\begin{equation}
\left(\begin{array}{cc}
\exp\left(\beta\Gamma\right) & 0\\
0 & \left(N-1+\exp\left(\beta\Gamma\right)\right)
\end{array}\right)+\beta\mathcal{B}\left(\begin{array}{cc}
\left(2\exp\left(\beta\Gamma\right)-1\right) & \left(\left(N-2\right)+2\exp\left(\beta\Gamma\right)\right)\\
1 & 1
\end{array}\right)\label{eq:Pertubration}
\end{equation}
\end{widetext}

Then 
\begin{equation}
E\cong\left(N-1+\exp\left(\beta\Gamma\right)\right)+\beta\mathcal{B}\label{eq:Energy}
\end{equation}
\begin{equation}
b\cong1\label{eq:Non_magnetic}
\end{equation}
\begin{equation}
a\cong\frac{\beta\mathcal{B}\left(N+2\exp\left(\beta\Gamma\right)-2\right)}{N-1}\label{eq:Magnetic}
\end{equation}
and 
\begin{equation}
M\cong\frac{2a}{N}=\frac{2\beta\mathcal{B}\left(N+2\exp\left(\beta\Gamma\right)-2\right)}{\sqrt{N}\left(N-1\right)}\label{eq:Magnetization-4}
\end{equation}
is the magnetization. This then means that at meanfield:
\begin{align}
\mathcal{B} & =M\mathcal{N}J\nonumber \\
\Rightarrow1 & =\frac{2\mathcal{\beta N}J\left(\left(N-2\right)+2\exp\left(\beta\Gamma\right)\right)}{N\left(N-1\right)}\label{eq:Solve}
\end{align}
Now we will assume that $\beta\mathcal{N}J\ll1$ so that $\left(N-2\right)\ll2\exp\left(\beta\Gamma\right)$
as such we obtain:
\begin{align}
1 & =\frac{4\mathcal{\beta N}J\exp\left(\beta\Gamma\right)}{N\left(N-1\right)}\nonumber \\
\beta\Gamma & =\log\left(\frac{N\left(N-1\right)}{4\beta\mathcal{N}J}\right)\nonumber \\
\lambda & =\frac{\log\left(\frac{N\left(N-1\right)}{4\beta\mathcal{N}J}\right)}{\beta J\mathcal{N}}\cong\frac{\log\left(N\left(N-1\right)/4\right)}{\beta J\mathcal{N}}\nonumber \\
\beta J\mathcal{N} & \cong\frac{\log\left(N\left(N-1\right)/4\right)}{\lambda}\label{eq:First_pass}
\end{align}
Iterating we get that: 
\begin{align}
1 & =2\frac{\log\left(N\left(N-1\right)/4\right)}{\lambda}\frac{\left(\left(N-2\right)+2\exp\left(\beta\Gamma\right)\right)}{N\left(N-1\right)}\nonumber \\
\beta\Gamma & \cong\frac{1}{2}\log\left(\frac{\lambda N\left(N-1\right)}{2\log\left(N\left(N-1\right)/4\right)}-N+2\right)\label{eq:Iteration}
\end{align}

\section{\protect\label{subsec:XXZ-model} Ferromagnetic Heisenberg model}

We consider the anisotropic 3D (quasi-1D) ferromagnetic Heisenberg
model with the following Hamiltonian: 
\begin{align}
H= & -\Gamma\sum_{i}\vec{\sigma}\left(i\right)\cdot\vec{\sigma}\left(i+\hat{z}\right)\nonumber \\
 & -J\sum_{\left\langle i,j\right\rangle ,j\neq i\pm\hat{z}}\vec{\sigma}\left(i\right)\cdot\vec{\sigma}\left(j\right)+B\sum_{i}\sigma_{i}^{z}\label{eq:Heisenberg_Model}
\end{align}
with $\mathcal{J}\gg J$. Now we consider the following meanfield
Hamiltonian (which happens to be the Heisenberg model in an external
field and as such solvable by BA techniques \citep{Zvyagin_2005,Takahashi_1999}):
\begin{equation}
H_{MF}=-\Gamma\sum_{i}\vec{\sigma}\left(i\right)\cdot\vec{\sigma}\left(i+1\right)-h\sum_{i}\sigma^{z}\left(i\right)\label{eq:Heisenberg_external_field}
\end{equation}
with 
\begin{equation}
h=\mathcal{N}J\mathcal{M}-B\label{eq:Condition}
\end{equation}
Where $\mathcal{M}$ will be chosen self consistently that is:
\begin{equation}
\mathcal{M}=\left\langle \sigma^{z}\left(i\right)\right\rangle _{H_{MF}}\label{eq:Self_consistent}
\end{equation}
We will now solve Eqs. (\ref{eq:Heisenberg_external_field}) and (\ref{eq:Self_consistent})
and find the critical transition temperature where $\mathcal{M}$
vanishes. This will be an extended meanfield treatment of the Hamiltonian
in Eq. (\ref{eq:Heisenberg_Model}). Indeed the magnetic susceptibility
$\chi$ can be used to determine phase boundaries. We have that for
the ferromagnetic Heisenberg model at temperature $T$ the susceptibility
is given by \citep{Takahashi_1999,Zvyagin_2005}:
\begin{equation}
\chi=\mathcal{J}^{-1}\left(\frac{1}{6}\left(\frac{\Gamma}{T}\right)^{2}+0.581\left(\frac{\Gamma}{T}\right)^{3/2}+0.68\left(\frac{\Gamma}{T}\right)\right)+...\label{eq:Approx}
\end{equation}
Now to determine the phase transition temperature between magnetic
and non-magnetic phase we use the relationship: 
\begin{align}
\chi h & =\mathcal{M}\nonumber \\
\chi\mathcal{N}J\mathcal{M} & =\mathcal{M}\nonumber \\
\chi\mathcal{N}J & =1.\label{eq:Phase_transition}
\end{align}
Where we have set $B=0$ to find the phase transition. As such for
the phase transition between magnetic and non-magnetic we write: 
\begin{equation}
\frac{\mathcal{N}J}{\Gamma}\left(\frac{1}{6}\left(\frac{\Gamma}{T}\right)^{2}+0.581\left(\frac{\Gamma}{T}\right)^{3/2}+0.68\left(\frac{\Gamma}{T}\right)\right)=1\label{eq:Transition-1}
\end{equation}
Now we have that 
\begin{align}
 & \frac{\mathcal{N}J}{\Gamma}\left(\frac{1}{6}\left(\frac{\Gamma}{T}\right)^{2}+0.581\left(\frac{\Gamma}{T}\right)^{3/2}+0.68\left(\frac{\Gamma}{T}\right)\right)\nonumber \\
 & \cong\frac{1}{6\lambda}\left(\frac{\Gamma}{T}\right)^{2}\label{eq:Simplification_leading_order}
\end{align}
This means that at first approximation for the critical temperature:
\begin{equation}
T_{0}\cong\Gamma\left(\frac{1}{6\lambda}\right)^{1/2}\label{eq:Iterate_zero}
\end{equation}
Now we write: 
\begin{align}
 & \frac{\mathcal{N}J}{\Gamma}\left(\frac{1}{6}\left(\frac{\Gamma}{T}\right)^{2}+0.581\left(\frac{\Gamma}{T}\right)^{3/2}+0.68\left(\frac{\Gamma}{T}\right)\right)\nonumber \\
 & \cong\frac{\mathcal{N}J}{\Gamma}\left(\frac{1}{6}\left(\frac{\Gamma}{T_{0}}\right)^{2}+0.581\left(\frac{\Gamma}{T_{0}}\right)^{3/2}+0.68\left(\frac{\Gamma}{T_{0}}\right)\right)\label{eq:Iterate}
\end{align}
As such the second approximation for the critical temperature gives:
\begin{equation}
T_{1}\cong\Gamma\sqrt{\frac{1}{6\lambda}\left[1-2.23\left(\frac{1}{\lambda}\right)^{1/4}-1.67\left(\frac{1}{\lambda}\right)^{1/2}\right]}\label{eq:Formula_final}
\end{equation}
We can continue to iteratively solve the problem for the critical
temperature more and more accurately using the formula (though Eq.
(\ref{eq:Formula_final}) is often sufficient): 
\begin{equation}
T_{n+1}\cong\Gamma\sqrt{\frac{1}{6\lambda}\left[1-\frac{1}{\lambda}\left[0.581\left(\frac{\Gamma}{T_{n}}\right)^{3/2}-0.68\left(\frac{\Gamma}{T_{n}}\right)\right]\right]}\label{eq:Iterative}
\end{equation}

\section{\protect\label{sec:Conclusions}Conclusions}

In this work we have extended the notion of the meanfield. We have
proposed a new class of meanfield models where 1D systems are the
meanfield problems to be solved. This combined with self consistency
relations may be used to efficiently solve for properties of anisotropic
3D or quasi-1D systems - where one direction is much more strongly
coupled then the other. In this work we have illustrated this idea
with the Ising and ferromagnetic Heisenberg models which are solvable
by transfer matrix and Bethe Ansatz techniques respectively and found
transition temperatures as preliminary results. We also studied a
more exotic model - the Potts model. We found significantly different
results then regular meanfield. Indeed we found magnetic phase transition
temperatures of $k_{B}T_{c}\sim\frac{\Gamma}{\log\left(\lambda\right)}$
for the Ising and Potts models and $k_{B}T_{c}\sim\frac{\Gamma}{\sqrt{\lambda}}$
for the ferromagnetic Heisenberg model, where as regular meanfield
predicts $k_{B}T_{c}\sim\Gamma$ for all three models. In future works
it would be of interest to extend these results to many other quasi-1D
systems systematically. Furthermore it would be of interest to study
cluster 1D meanfields, where a cluster of 1D systems is chosen as
the meanfield system to be solved. The accuracy of the calculation
is systematically improved with cluster size in cluster 1D meanfields.

\textbf{Acknowledgements:} The author would like to thank Natan Andrei
for useful discussions.

\appendix

\section{\protect\label{sec:Onsanger-solution}Onsager solution}

We note that the Onsager relation for $d=2$ ($\mathcal{N}=2$) is
that the critical temperature is given by \citep{Pathria_2011,Onsager_1944}:
\begin{align}
\sinh\left(2\beta J\right)\sinh\left(2\beta\Gamma\right) & =1\nonumber \\
\exp\left(2\beta\Gamma\right)\cdot\beta J & \cong1\label{eq:Onsanger}
\end{align}
This is the exact same relationship as in Eq. (\ref{eq:Magnetism})
for $\mathcal{N}=1$, which means that: 
\begin{equation}
\beta J=\frac{1}{2\lambda\log\left(2\lambda\right)}+...\label{eq:Solution-1-1}
\end{equation}
This means the meanfield is accurate within a factor of two while
the regular meanfield is off by $\sim\log\left(2\lambda\right)$ for
large $\lambda$ showing significant improvement of our approach over
regular meanfield in the limit of strong anisotropy or nearly 1D systems.

\section{\protect\label{sec:-state-Potts-model}$N$-state Potts model in
an external Field at unit temperature}
\begin{widetext}
Now we look for the biggest eigenvalue using the ansatz in Eq. (\ref{eq:Diagonal}).
This means that: 
\begin{equation}
\left(\begin{array}{cc}
\left(\exp\left(J+2B\right)-\exp\left(B\right)+1\right) & \left(\left(N-2\right)\left[\exp\left(B\right)-1\right]+\exp\left(J+2B\right)-\exp\left(J\right)\right)\\
\left[\exp\left(B\right)-1\right] & \left(N-2+\exp\left(B\right)+\exp\left(J\right)\right)
\end{array}\right)\left(\begin{array}{c}
a\\
b
\end{array}\right)=E\left(\begin{array}{c}
a\\
b
\end{array}\right)\label{eq:Eigenvalue}
\end{equation}
This means that 
\begin{equation}
\det\left[\begin{array}{cc}
\left(\exp\left(J+2B\right)-\exp\left(B\right)+1-E\right) & \left(\left(N-2\right)\left[\exp\left(B\right)-1\right]+\exp\left(J+2B\right)-\exp\left(J\right)\right)\\
\left[\exp\left(B\right)-1\right] & \left(N-2+\exp\left(B\right)+\exp\left(J\right)-E\right)
\end{array}\right]=0\label{eq:Zero}
\end{equation}
As such: 
\begin{equation}
E^{2}-E\left[\exp\left(J+2B\right)+N-1+\exp\left(J\right)\right]+\exp\left(2J+2B\right)+\left[N-1\right]\left[\exp\left(J+2B\right)-\exp\left(2B\right)+\exp\left(B\right)\right]=0\label{eq:Characteristic}
\end{equation}
As such (since we want the biggest eigenvalue): 
\begin{equation}
E_{+}=\frac{\left[\exp\left(J+2B\right)+N+\exp\left(J\right)-1\right]+\sqrt{\Delta\left(J,B,N\right)}}{2}\label{eq:Enrgy}
\end{equation}
Where 
\begin{equation}
\Delta\left(J,B,N\right)=\left[\exp\left(J+2B\right)-N+1+\exp\left(J\right)\right]^{2}-4\left[\exp\left(2J+2B\right)+\left[N-1\right]\left[\exp\left(B\right)-\exp\left(2B\right)-\exp\left(J\right)\right]\right]\label{eq:Definition}
\end{equation}
\end{widetext}

Now we have that 
\begin{align}
\frac{a}{b} & =\frac{E_{+}-N+2-\exp\left(B\right)-\exp\left(J\right)}{\exp\left(B\right)-1}\nonumber \\
 & =\frac{\exp\left(J+2B\right)+\sqrt{\Delta\left(J,B,N\right)}}{2\left(\exp\left(B\right)-1\right)}-\frac{1}{2}\label{eq:ratio}
\end{align}
Now we have that the normalization of the vector is 
\begin{equation}
\left\Vert a\left|1\right\rangle +b\left|V\right\rangle \right\Vert =Nb^{2}+a^{2}+2ab\label{eq:Normalization}
\end{equation}
Therefore the magnetization is given by:
\begin{equation}
M=\frac{a^{2}+2ab}{Nb^{2}+a^{2}+2ab}\label{eq:Magnetization-2}
\end{equation}

\end{document}